\documentclass[twocolumn,showpacs,aps,pre,groupedaddress,amssymb,amsmath]{revtex4}

\usepackage{graphicx}% Include figure files
\usepackage{longtable}% Include figure files

\begin{document}

\title{Flow transitions in two-dimensional foams}

\author{Christopher Gilbreth}
\author{Scott Sullivan}
\author{Michael Dennin}
\affiliation{Department of Physics and Astronomy, University of
California at Irvine, Irvine, California 92697-4575}

\date{\today}

\begin{abstract}
For sufficiently slow rates of strain, flowing foam can exhibit
inhomogeneous flows. The nature of these flows is an area of
active study in both two-dimensional model foams and three
dimensional foam. Recent work in three-dimensional foam has
identified three distinct regimes of flow [S. Rodts, J. C. Baudez,
and P. Coussot, Europhys. Lett. {\bf 69}, 636 (2005)]. Two of
these regimes are identified with continuum behavior (full flow
and shear-banding), and the third regime is identified as a
discrete regime exhibiting extreme localization. In this paper,
the discrete regime is studied in more detail using a model two
dimensional foam: a bubble raft. We characterize the behavior of
the bubble raft subjected to a constant rate of strain as a
function of time, system size, and applied rate of strain. We
observe localized flow that is consistent with the coexistence of
a power-law fluid with rigid body rotation. As a function of
applied rate of strain, there is a transition from a continuum
description of the flow to discrete flow when the thickness of the
flow region is approximately 10 bubbles. This occurs at an applied
rotation rate of approximately $0.07\ {\rm s^{-1}}$.

\end{abstract}

%\keywords{}

\maketitle

\section{Introduction}

Nonlinear viscoelastic materials are observed to exhibit
inhomogeneous flows. One source of inhomogeneous flow is
geometric. When subjected to certain boundary conditions, an
inhomogeneous stress will be applied to the fluid, resulting in an
inhomogeneous flow. A classic example is a yield stress fluid in a
Couette geometry (flow between concentric cylinders). A yield
stress fluid only flows if the stress is above a critical stress,
and in a Couette geometry, the stress decreases as a function of
the radial distance from the inner cylinder. Therefore, it is
possible to generate flow close to the inner cylinder where the
stress is above the yield stress, and at the same time, there will
be no flow beyond a critical radius. The critical radius
corresponds to the point at which the stress has decreased below
the yield stress \cite{BOOKS}. Though inhomogeneous flows are not
a new phenomenon, it has been challenging to directly measure them
due to the inherent opacity of most complex fluids, such as foams,
granular matter, colloids, emulsions, pastes, etc.. Direct
measurements of such flows has focused on two-dimensional systems
\cite{DTM01,LCD04} (for which the entire material is observable)
and non-optical techniques in three dimensional systems, such as
MRI techniques \cite{CRBMGH02}. A surprising element of these
direct studies of velocity profiles is the observation of
inhomogeneous flow that can not be understood in terms of
the geometry and the resulting stress distribution.

Direct measurement of velocity profiles has covered a range of
diverse systems. These include worm-like micelles
\cite{SCMM03,LAD06}, lyotropic lamellar phases
\cite{SBMC03b,SMC03c}, granular matter
\cite{HBV99,LBLG00,MDKENJ00}, slurries and pastes
\cite{CRBMGH02,HOBRCD05}, and foams
\cite{GD99,DTM01,LTD02,RCVH03,LCD04,RBC05}. In most of these
studies, inhomogeneous flows take the form of {\it flow
localization} or {\it shear localization}. This refers to the fact
that the system divides into two spatial regions: a flowing region
and a stationary (or solid-like) region. The breadth of systems
exhibiting this general behavior naturally leads to questions of
universality of the observed flow localization. In this regard,
the proposed jamming transition \cite{LN98} provides a natural
context for considering flow localization. Essentially, the
jamming transition refers to the transition from solid to
fluid-like behavior (or vice-versa) in a system as a function of
density, temperature, or externally applied stress. Of particular
interest is the correspondence between transitions due to a
critical stress and the more familiar transitions (such as the
glass transition) as a function of temperature and density
\cite{ITPJamming}. If one is to consider shear localization in
this context, it is natural to divide the behavior into two
categories: continuous and discontinuous transitions.

The division into continuous and discontinuous transitions is
based on the expected analogy to phase transitions. In the case of
jamming transitions as a function of applied stress, the relevant
variable is the rate of strain. In continuous transitions, the
rate of strain is continuous as the system makes the transition
from flow to no-flow. For this class of behavior, the velocity as
a function of position in the system is often well-fit by an
exponential \cite{LBLG00,DTM01}. Such continuous behavior was the
first class of inhomogeneous flows that was identified as
potentially universal across different systems, having been
observed in both granular and foam systems. Though not central to
the work in this paper, it is worth noting that in granular
matter, the velocity profile has been attributed to a spatial
variation in the density that develops during flow \cite{LBLG00}.
This results in a spatial variation of the viscosity that is the
source of the shear localization. For foam, exponential profiles
were observed in foam confined in a Hele-Shaw cell \cite{DTM01}.
However, experiments \cite{wkd05a} and simulation \cite{WJH06}
suggest that this type of continuous transition is caused by drag
due to the confining plates.

More recently, experiments have identified examples of
discontinuous transitions. Such flows have been observed in
various slurries and pastes \cite{CRBMGH02,HOBRCD05}, including
three-dimensional foam \cite{RBC05}, in two-dimensional model
foams known as bubble rafts \cite{LCD04}, and in worm-like
micelles \cite{SCMM03} and lyotropic lamellar systems
\cite{SBMC03b}. In these flows, the rate of strain is
discontinuous across the system, often at a transition from shear
flow to rigid body type behavior. Discontinuous flows have not yet
been observed in simulations, though a number of simulations
exhibit continuous flow localization
\cite{KD03,VBBB03,XOK05a,XOK05b,WJH06}.

In the context of a jamming transition, a discontinuous transition
represents the natural analog to a first order phase transition.
Therefore, discontinuous transitions represent an important class
of transitions if one is to fully understand the implications of
the jamming paradigm. Furthermore, within the context of
rheological models, these discontinuous flows require non-standard
constitutive relations. Essentially all the standard stress-strain
relations used to describe complex fluids assume a continuous
dependence of the stress on the rate of strain, and hence, the
rate of strain must be continuous by definition for any transition
from fluid to solid-like behavior.

The above discussion assumes that the materials are well described
by a continuum model, either in the context of a specific
constitutive relation or the jamming transition. For the case of
flow in a three-dimensional foam, there has been a detailed study
of the flow behavior that combines standard rheological measures
of stress with direct measurement of velocity profiles using a
Couette geometry \cite{RBC05}. For these studies, a large system
size was used that allowed for the categorization of the flow into
two classes. When the width of the flowing region is above a
critical value, a single constitutive relation that is based on
the existence of a critical rate of strain is used to describe the
flow of the foam over a wide range of rates of strain. This is to
be expected in the continuum limit. When the width of the flowing
region is less than a critical value, a different type of behavior
is observed. This has been called the ``discrete flow'' regime.

The discrete regime was observed to occur when the flow is
localized to a spatial width that is less than approximately 25
bubble diameters. This corresponded to a critical rotation rate
for the driving cylinder (in this case the inner cylinder) on the
order of $0.3\ {\rm s^{-1}}$ \cite{RBC05}. For the discrete
regime, there does not exist a single continuum model that
describes all of the flow curves. For example, when the system is
driven by rotating the inner cylinder with a constant rotation
rate, the torque as a function of the applied rotation rate does
not follow a well-defined curve.

Previous measurements on a bubble raft suggest that the discrete
regime occurs in two dimensional foam as well \cite{LCD04}. A
bubble raft consists of a single layer of bubbles floating on the
water surface \cite{BL49,AK79}. Studies of the flow behavior of a
bubble raft using a Couette geometry demonstrated the coexistence
of flowing and non-flowing regions with a rate of
strain discontinuity between the two regions \cite{LCD04}. In
these studies, constant rotation of the outer cylinder was used to
generate the flow. Two different external rotation rates were
studied, and in both cases, the velocity profile in the flowing
region was well described by a power-law fluid. However, different
power-law fluid models were required for each case. Because the
total system size was only 25 - 30 bubbles across, this is
consistent with the expectation of a discrete flow regime. The
possible connection between these flows and the observed discrete
regime in three dimensional fluids provided the motivation for the
study reported on in this paper.

In this paper, we report on a more complete study of the flow in
the small system size bubble raft to elucidate {\it both} the
nature of the solid to fluid transition (i.e. is it continuous or
discontinuous) {\it and} to determine if there is a transition
from the continuum limit to discrete flow. We focused on low rates
of strain and considered two different system sizes. We use a foam
confined between two cylinders, and apply a constant rate of
rotation of the outer cylinder. We confirm the existence of a
transition between a discrete flow regime and a continuum limit as
a function of the rotation rate of the outer cylinder. In both
cases, the flow can be described as the coexistence of a power-law
type fluid and a rigid body, with a discontinuity in the rate of
strain. However, in the discrete regime, one can not use a single,
consistent power-law model for all rotation rates. The rest of the
paper is organized as follows. Section II describes the
experimental setup and techniques. Section III discusses the
models used to describe the data, and Sec. IV presents the
results. Section V is a discussion of the results.

\section{Experimental methods}

The experimental system consisted of a standard bubble raft
\cite{AK79} in a Couette geometry (two concentric cylinders). The
system was driven by rotating the outer cylinder at a constant
angular speed. Stress measurements were made using the inner
cylinder, which was suspended on a torsion pendulum. The studies
used two values of the outer radius ($R$): $7\ {\rm cm}$ and $9\
{\rm cm}$. The inner cylinder had a fixed radius of $r_i = 2.85\
{\rm cm}$. The bubble raft was produced by flowing regulated
nitrogen gas through a hypodermic needle into a homogeneous
solution of 80\% by volume deionized water, 15\% by volume
glycerine, and 5.0\% by volume Miracle Bubbles (Imperial Toy
Corp.). The bubble size was dependent on the nitrogen flow rate,
which was varied using a needle valve.  A random distribution of
bubble sizes was used, with an average radius of $1.7\ {\rm mm}$
and 15-22 bubbles across, depending on the outer radius. In
previous versions of this system, the bubbles were generated
separately and transferred into the apparatus. For these
experiments, the bubbles were generated directly in the Couette
apparatus. For each setting of the outer radius, essentially the
same set of bubbles were used. Occasionally, near the end of a
run, some bubbles would pop. Data was only used up to the point
the first bubbles were observed to pop, and the bubble raft was
filled in before the next run. For additional details of the
apparatus, see Ref.~\cite{app}.

As mentioned, the system was driven by rotating the outer cylinder
at a constant angular velocity, $\Omega$. The range of angular
speeds used was $0.01\ {\rm s^{-1}} \leq \Omega \leq 0.35\ {\rm
s^{-1}}$. The first layer of bubbles at either boundary was not
observed to slip relative to the boundary. At the outer boundary,
this was due to the curvature of the boundary. At the inner
boundary the first layer of bubbles was held in place with metal
fins attached to the boundary. Due to the finite size of the
bubbles, this resulted in an effective inner radius on the order
of $3.1$ to $3.2\ {\rm cm}$, depending on the details of the
system.

As the experiments focus on the average velocity profile and the
corresponding rate of strain (or shear rate) as a function of
radial position, it is useful to review what is expected for
Newtonian fluids in a Couette geometry. Due to the cylindrical
geometry, the shear rate is not uniform across the system and is
given by $\dot{\gamma}(r) = r
\frac{d}{dr}\frac{v_{\theta}(r)}{r}$. Here $v_{\theta}$ is the
azimuthal velocity of the bubbles. Because we rotate the outer
cylinder, it is useful to normalize the velocity by the expected
velocity for rigid body rotation $V(r) = \Omega r$, where $r$ is
the radial position of interest. Therefore, we will often refer to
the scaled azimuthal velocity $v(r)= v_{\theta}(r)/\Omega r$. It
should be noted that because we rotate the outer cylinder, the
bubbles in the outer portion always move.  Therefore, with our
setup, what distinguishes the ``flowing'' region from the
``solid'' region is the type of motion. The solid region exhibits
rigid-body rotation, and has $v(r) = 1$. This is in contrast to
many experiments in which the inner cylinder drives the system and
``solid'' behavior corresponds to a zero velocity for
bubbles in the outer portion of the system.

To measure the velocity, roughly one third of the trough was
digitally recorded using a frame grabber. The time interval
between images was selected so that the fastest moving bubbles
could be accurately tracked from frame to frame. The radial
coordinate is divided into bins of width $0.15\ \rm{cm}$ and
$0.18\ \rm{cm}$ for the $7\ \rm{cm}$ and $9\ \rm{cm}$ data,
respectively. This represents roughly an average bubble radius per
bin. The details of the velocity measurements are given in
Ref~\cite{LCD04}.

The other bubble motion of interest is the nonlinear bubble
rearrangements, referred to as T1 events. A T1 event is a neighbor
switching event that involves four bubbles. Two bubbles that are
initially neighbors lose contact, and two bubbles that were not
neighbors become neighbors. In contrast to the automated velocity
measurements, T1 events were measured by stepping the digitized
images one frame at a time and visually searching for the location
and time at which T1 events occurred. Due to the associated
motions of the other neighboring bubbles, T1 events are relatively
easy to detect by hand, and for all reports of T1 events, two
researchers independently tracked the events. It should be noted
that the velocity can be determined automatically even if not
every bubble is tracked. However, to reliably track T1 events
requires monitoring every bubble. This is the reason for the need
for identifying T1 events by hand. Currently, we are working on
improving this apparatus so tracking T1 events is automated (as it
has been done in other systems in our lab \cite{wkd05a}).

\section{models}

The data reported in Sec.~IV will be discussed in the context of
two standard continuum models for non-Newtonian fluids:
Herschel-Bulkley model and power law fluid model \cite{BAH77}. The
choice of these models allows us to test both the continuity of
the transition between the fluid and solid behavior and the
transition from continuum to discrete behavior. By construction,
the Herschel-Bulkley model is continuous in the rate of strain,
and the power law model allows for a rate of strain discontinuity.
We will show that the power law model, with a rate of strain
discontinuity, is the most consistent with the data. Once this is
established, the power law model provides an effective method of
characterizing the transition from the continuum limit to the
discrete flow regime. In this section, we review the main elements
of each model that are used to analyze the data. The details of
the derivations are left to the Appendix, as indicated.

The key element of the Herschel-Bulkley model is the yield stress,
$\tau_0$. If the stress is below the yield stress, the material
acts as a solid. For stresses above the yield stress, the material
obeys the following constitutive relation for the average stress
$\sigma$ as a function the of rate of strain $\dot{\gamma}$:
\begin{equation} \label{eqn:HB1}
\sigma(\dot{\gamma}) = \tau_0 + \mu \dot{\gamma}^n
\end{equation}
\cite{BAH77}. As one can see, this model is explicitly continuous
in the rate of strain. Also, it is important to point out that it
provides a good fit to previous measurements of stress versus rate
of strain in bubble rafts. Some typical values for a bubble raft
are $\tau_0 = 0.8 \pm 0.1\ {\rm mN/m}$ and $n = 0.33$
\cite{PD03,LCD04}. Second, the model has a built in physical
mechanism for the transition from solid-like behavior to flow: the
yield stress.

The second model was selected based on previous velocity profiles
\cite{LCD04} and MRI measurements of velocity profiles in three
dimensional foam \cite{RBC05}. These experiments suggest that the
system is best described by a power-law fluid coexisting with the
elastic solid regime. For a generic power-law fluid,
\begin{equation} \label{eqn:PL2}
\sigma(\dot{\gamma}) = \mu \dot{\gamma}^n,
\end{equation}
In this case, there is no specification of where the fluid-solid
transition will occur. To account for this, a modified power-law
fluid that explicitly includes a critical rate of strain
($\dot{\gamma}_c$) was introduced in Ref.~\cite{RBC05}:
\begin{equation} \label{eqn:PL1}
\sigma(\dot{\gamma}) = \mu (\dot{\gamma}/\dot{\gamma}_c)^n\ {\rm
for}\ \dot{\gamma} > \dot{\gamma}_c,
\end{equation}
for $\dot{\gamma} < \dot{\gamma}_c$, $\dot{\gamma} = 0$. Notice,
this is fundamentally different from the Herschel-Bulkley model in
which $r_c$ is set by a critical stress. For this power-law model,
$r_c$ is determined by $\dot{\gamma}_c$, and there is a built in
discontinuity in $\dot{\gamma}$ at $r_c$. Localized flow occurs
when there is a coexistence of a flowing region for $r < r_c$
(where $\dot{\gamma}
>\dot{\gamma}_c$) and a ``solid'' region for $r > r_c$ (where
$\dot{\gamma} = 0$).

Independent of the model, the stress in a Couette geometry has the
form $\sigma(r) = C/r^2$, where the constant $C$ is determined by
the boundary conditions (either at the inner cylinder or $r_c$,
whichever is more convenient). Combining this relation with the
particular constitutive model, allows one to solve for $v(r)
\equiv v_{\theta}/\Omega r$. This in turn can be fit to the
velocity data as a test for each model. For the Herschel-Bulkley
model, we get for $r \leq r_c$,
\begin{equation} \label{eqn:HB3a}
v(r) = \frac{1}{N}
   \int_{r_i}^{r} \frac{1}{\rho}\Big(\Big(\frac{r_c}{\rho}\Big)^{2} - 1 \Big)^{1/n}
   d\,\rho
\end{equation}
where
\begin{displaymath}
N = \int_{r_i}^{r_c}
\frac{1}{\rho}\Big(\Big(\frac{r_c}{\rho}\Big)^{2} - 1 \Big)^{1/n}
   d\,\rho.
\end{displaymath}
For $r \geq r_c$, the solid body behavior gives $v = 1$; adjoining
the solutions for $r \leq r_c$ and $r \geq r_c$ yields a smooth
curve where the fit parameters are the exponent $n$, $r_i$ and
$r_c$. For the power-law case, we find
\begin{equation} \label{eqn:PL3}
v(r) = \frac{A}{r^{2/n}} - B.
\end{equation}
Applying the same boundary conditions as before, $A = (r_i
r_c)^{2/n}/(r_i^{2/n}-r_c^{2/n})$ and $B =
r_c^{2/n}/(r_i^{2/n}-r_c^{2/n})$. In this case, because of the
rate of strain discontinuity, a smooth continuation of the
power-law solution and the solid body curve does not exist.
Therefore, the data in the range $0.25 < v < 0.95$ are fit to
Eq.~\ref{eqn:PL3} with $n$, $r_i$ and $r_c$ as fit parameters.
Equivalently, $r_c$ can be computed as the intersection the line
$v = 1$ ($v_{\theta} = \Omega r$) with a fit to Eq.~\ref{eqn:PL3}
with $A$, $B$ and $n$ as free parameters. Notice, $\dot{\gamma}_c$
is not a fit parameter, as it does not appear in the solution for
the power-law model. To determine $\dot{\gamma}_c$, one uses the
resulting fit and computes $\dot{\gamma}(r_c)$.

In terms of testing the models for their applicability, it is
important to note a few similarities and differences. In both
models the three main fit parameters are $n$, $r_i$ and $r_c$. The
determination of $r_c$ and $r_i$ provides a consistency check, as
both (within a few percent) are relatively easy to determine by
visual inspection of the data, independent of the selected model.
Therefore, one does not expect to be able to distinguish the
models on this basis. On the other hand, $n$ has definite physical
meaning. It gives the scaling of the stress (or the effective
viscosity) as a function of rate of strain. For $n < 1$, the
material acts as a shear thinning material. For $n > 1$, the
material acts as a shear thickening material. All of the
measurements of stress versus rate of strain suggest that $n < 1$,
so this provides a physical test that the models must meet to be
considered applicable. Another useful test of the models is the
behavior of the rate of strain at $r_c$. In this case, the models
predict very different behavior by construction. For the
Herschel-Bulkley model the rate of strain is continuous at $r_c$,
while for the power law model there is an explicit discontinuity.
Here, because of the discrete nature of the data, it is difficult
to determine the continuity of the data by visual inspection. As
we will show, other aspects of the fit clarify the nature of the
transition.

There is an additional test of the applicability of the
Herschel-Bulkley model: the behavior of $r_c$ as a function of the
external rotation rate $\Omega$. Because the critical radius is
set by the yield stress, $\sigma(r_c) = \tau_0$. From this, we can
write $\sigma(r) = \tau_0 r_c^2 / r^2$. Then as the rotation rate
approaches zero, the rate of strain in Eq.~(\ref{eqn:HB1}) does as
well, so $\sigma(r_i)$ approaches $\tau_0$ and thus $r_c$
approaches $r_i$.

The final question related to the choice of models is the use of
these models in a discrete regime. By definition, both the
Herschel-Bulkley model and the critical rate of strain power law
model discussed here are continuum models. The key element of each
is a single yield stress or critical rate of strain, respectively,
that describes the properties of the foam in a continuum limit. It
is difficult to modify the Herschel-Bulkley model to also describe
a situation in which there is not a well-defined yield stress.
However, if we consider the original general power law model, it
is useful as an ad hoc model in the case of discrete flow. As we
saw, the solution for $v(r)$ in the case of a power-law fluid does
not depend on $\dot{\gamma}_c$. Therefore, one would expect the
fitting procedure to work equally well whether the system is
exhibiting continuum or discrete behavior. However, the {\it
results} can be used to distinguish between the two regimes based
on the behavior of $\dot{\gamma}_c = \dot{\gamma}(r_c)$ and $n$.
For a system in a well-defined continuum limit, these are expected
to be material properties that are independent of the external
rotation rate $\Omega$. If one observes a strong dependence of
these parameters on $\Omega$, this would represent a break down of
the continuum limit.

In summary, by considering two standard models for complex fluids
(power-law fluid and Herschel-Bulkley) we will be able to test two
features of the flow. First, we will be able to distinguish
between transitions that are continuous or discontinuous in the
rate of strain. Second, we will be able to test for a transition
from the continuum limit to discrete flow.

\section{results}

Figure 1 illustrates the fitting procedure for the two models for
a rotation rate of $0.09\ {\rm s^{-1}}$ and outer radius $R = 9\
{\rm cm}$. The power law fit is illustrated in Fig.~1a, where the
horizontal line indicates the rigid body rotation. The fit is
continued past the critical radius to highlight the fact that this
model gives a non-zero critical rate of strain. This represents a discontinuous
transition. In contrast, the fit in Fig. 1b is for the
Herschel-Bulkley, and it is continuous by construction. From these
fits alone, it is difficult to distinguish between these two
models, especially at this rotation rate. Therefore, one needs to
consider more carefully the parameters for the different models,
as they are determined from the fits.

\begin{figure}
\includegraphics[width=8.5cm]{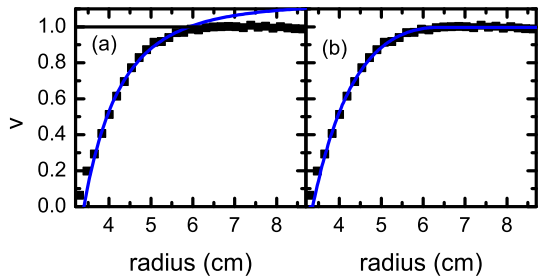}
\caption{(color online) (a) A velocity profile (with $r_c = 5.9\ \rm{cm}$) as a function of
radial position for the system with outer radius of 9 cm and
rotation rate of $0.09\ \rm{ s^{-1}}$ (black symbols), with power
law fit where $n = 0.515$ (blue line). (b) Same velocity profile (black symbols),
with numeric fit to Herschel-Bulkley model where $n = 1.056$ (blue line).}
\label{profiles}
\end{figure}

Figure 2 provides a summary of the behavior of the critical radius
as a function of system size and rotation rate. For comparison, we
show both the calculation using the numerical fit to the
Herschel-Bulkley model (open symbols) and the fit to a power
law/rigid body coexistence model (closed symbols). The results for
$r_c$ based on each method are in reasonable agreement. This
provided an important consistency check on both methods. It should
be noted that once $r_c \approx R$, the various methods of
determining $r_c$ break down. To indicate where this occurs, we
still plot $r_c$, but define it to be equivalent to $R$. In other
words, $r_c = R$ is the condition that the entire sample is
flowing.

\begin{figure}
\includegraphics[width=8.5cm]{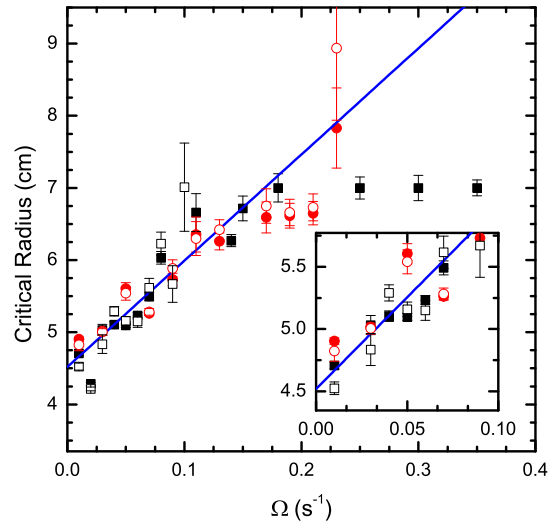}
\caption{(color online) Comparison of the measured critical radius
as a function of rotation rate for outer radii of 7 cm (black
squares) and 9 cm (red circles). The open symbols correspond to
calculation of $r_c$ using a Herschel-Bulkley model, and the solid
symbols correspond to calculating $r_c$ assuming a power-law fluid
coexisting with a solid-like state.} \label{critical radius}
\end{figure}

Three important features are highlighted by this figure. First,
the critical radius is roughly linear in the applied rotation
rate. (The straight line in the figure is a guide to the eye.)
Second, for rotation rates less than $0.15\ {\rm s^{-1}}$,
changing the system size does not alter the location of the
coexistence between the two regions. Once the critical radius is
$r_c = 7\ {\rm cm}$, in the $R = 9\ {\rm cm}$ system, the critical
radius continues to increase with rotation rate. Finally, the
critical radius approaches $r = 4.5\ {\rm cm}$ as $\Omega$ goes to
zero. The fact that $r_c \rightarrow 4.5\ {\rm cm}$ as $\Omega
\rightarrow 0\ {\rm s^{-1}}$ is the first evidence for the
breakdown of the Herschel-Bulkley model, as discussed in Sec. III.
As a further test of the behavior of $r_c$, we have considered two
slower rates of strain: $1.0 \times 10^{-3}\ {\rm s^{-1}}$ and
$3.0 \times 10^{-3}\ {\rm s^{-1}}$. Due to the finite lifetime of
the bubbles, the results for the velocity profiles were noisier at
these extremely slow rates of strain than the data studied in
detail in this paper. However, these profiles clearly exhibited a
$r_c$ that was close to but greater than $4.5\ {\rm cm}$,
consistent with the results reported in Fig.~2.

\begin{figure}
\includegraphics[width=8.5cm]{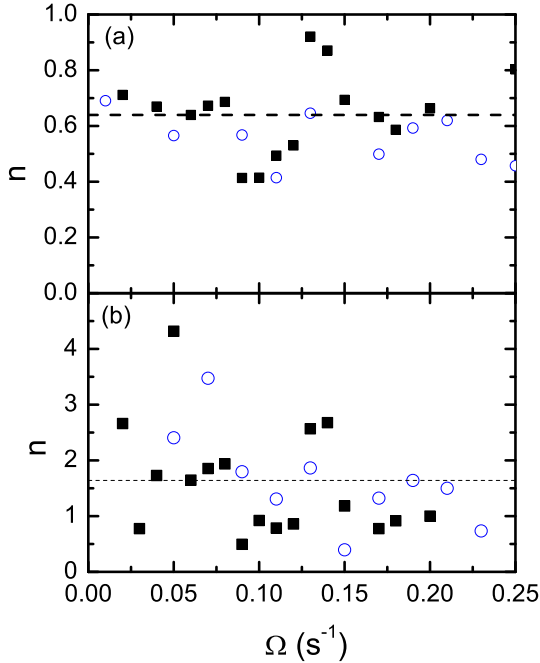}
\caption{(color online) (a) The exponent $n$ obtained from fitting
the velocity curves to a power-law fluid coexisting with a
solid-like state versus the external rotation rate for $R = 7\
{\rm cm}$ (black squares) and $R = 9\ {\rm cm}$ (blue circles).
(b) The exponent $n$ obtained from fitting the velocity curves to
a Herschel-Bulkley model versus the external rotation rate for $R
= 7\ {\rm cm}$ (black squares) and $R = 9\ {\rm cm}$ (blue
circles).} \label{scaled vel}
\end{figure}

As discussed, the exponent $n$ in both the Herschel-Bulkley model
and the power-law model has physical significance. Figure 3
summarizes the values of $n$ obtained in the various fits to the
velocity profiles. Figure 3a is the results for the power-law
model, with the $R = 7\ {\rm cm}$ and $R = 9\ {\rm cm}$ indicated
by squares and circles, respectively. Figure 3b is the results for
the Herschel-Bulkley model. The dashed line in each figure
represents the median value for the exponents. There are two
striking features in Fig.~3. First, neither fit provides a
completely consistent value for $n$. However, the variation in the
fits for the power-law model is smaller than the case of the
Herschel-Bulkley model. Second, for the power-law fits, one
consistently finds $n < 1$, but for the Herschel-Bulkley model,
one consistently finds $n > 1$ (though there are some isolated
cases for which $n < 1$).

\begin{figure}
\includegraphics[width=8.5cm]{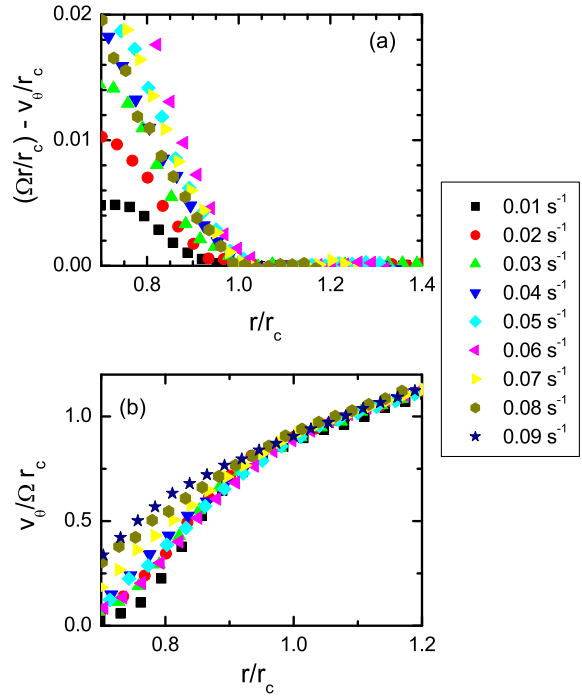}
\caption{(color online) (a) Plot of the scaled velocity profiles
$\tilde{v} = (\Omega r/r_c) - v_{\theta}/r_c$ versus $r/r_c$. (b)
Plot of $v_{\theta}/\Omega r_c$ versus $r/r_c$. Rates of strain
are indicated in the legend.} \label{exponent}
\end{figure}

As the combination of Figs.~2 and 3 effectively rule out a
Herschel-Bulkley model, it is necessary to further probe the
applicability of the power-law model. This allows us to confirm
whether or not the transition is truly discontinuous in the rate
of strain and to look a transition from continuum to discrete
flow. One approach is to consider the scaling of the velocity
profiles. In this regard, there are two different scalings that
are particularly useful. These are presented in Fig.~4, and these
represent the central measurement of both the discontinuity in the
transition and the transition from continuum to discrete flow.
First, it is useful to directly test for the existence of a
consistent critical rate of strain in the context of the power-law
model. To do this, we follow the scaling arguments presented in
Ref.~\cite{RBC05}. Because we are rotating the outer cylinder, we
first subtract out the rigid body rotation. This gives a new
transformed velocity: $\tilde{v} = (\Omega r/r_c) -
v_{\theta}/r_c$. In this form, $\tilde{v} = 0$ at $r = r_c$.
Therefore, rewriting the rate of strain in the form
$\dot{\gamma}(r) = r \frac{d}{dr}\frac{v_{\theta}(r)}{r} =
\frac{dv_{\theta}}{dr} - \frac{v_{\theta}}{r}$, we see that the
slope of $\tilde{v}$ at $r = r_c$ is the critical rate of strain
$\dot{\gamma}_c$. Therefore, if there is a single $\dot{\gamma}_c$
for the material, then $\tilde{v}$ as a function of $r/r_c$ will
collapse onto a single curve.

Figure~4a presents $\tilde{v}$ as a function of $r/r_c$. This
illustrates two features. First, for sufficiently high rates of
strain, the data suggests a single value of $\dot{\gamma}_c$.
However, the scaling breaks down below a critical value of the
external rotation rate. Second, this form of the velocity
highlights the discontinuity in the slope at the critical radius.

Given the apparent dependence of $\dot{\gamma}_c$ on $\Omega$ for
slow rates of strain, it is useful to consider an alternative
scaling of the data, given in Fig.~4b. Here we plot
$v_{\theta}/\Omega r_c$ as a function of $r/r_c$. For this case,
we do not subtract the rigid body behavior, which is apparent as
the linear regime for $r/r_c > 1$. This was done to confirm the
consistent rigid body behavior for all rotation rates. This
scaling allows us to focus on the data for the slowest rotation
rates, which did not scale in Fig.~4a. In this case, we observe a
collapse of the data for slow rates of strain, indicating a
dependence of $\dot{\gamma}_c$ on $\Omega$.

\begin{figure}
\includegraphics[width=8.5cm]{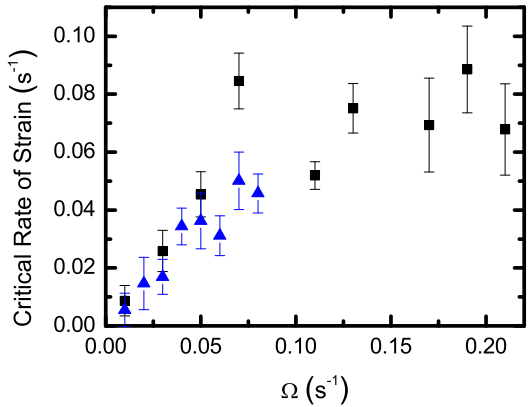}
\caption{(color online) Critical rate of strain as a function of
external rotation rate for $R = 7\ {\rm cm}$ (blue triangles) and
$R = 9\ {\rm cm}$ (black squares).} \label{critical rate of
strain}
\end{figure}

The final measurement of the discrete/continuum flow transition is
the direct measurement of $\dot{\gamma}_c$ as a function of
$\Omega$. This is plotted in Fig.~5. In this case, we fit the data
for $\tilde{v}$ near $r_c$. Two features of are illustrated by the
data. First, for rotation rates below $\Omega = 0.07 \pm 0.02\
{\rm s^{-1}}$, the critical rates of strain are dependent on the
external rotation rate, but independent of the system size.
Interestingly, the behavior is consistent with a linear dependence
on $\Omega$. This is strong evidence for the break down of any
continuum description of the flow. Second, we observe a cross over
to a regime in which $\dot{\gamma}_c$ is independent of $\Omega$.
This occurs for value of the external rotation above $\Omega =
0.07 \pm 0.02\ {\rm s^{-1}}$ and is consistent with the scaling of
$\tilde{v}$ presented in Fig.~4a. From this data, we find
$\dot{\gamma}_c = 0.07 \pm 0.01\ {\rm s^{-1}}$, where the error
represents the standard deviation of the measured values of the
critical rate of strain, for our bubble raft.

\begin{figure}
\includegraphics[width=8.5cm]{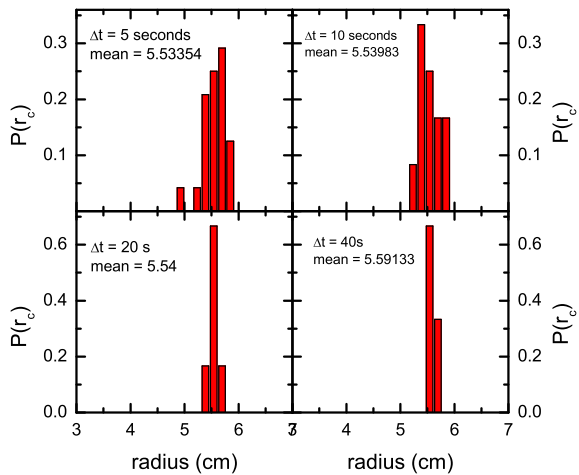}
\caption{Probability distribution of measuring a particular value
of $r_c$ based on averaging the velocity over finite time
intervals. Four different time intervals are illustrated for a
rotation rate of $0.07\ {\rm s^{-1}}$ with $R = 7\ {\rm cm}$. The
time intervals and mean for each distribution are indicated in the
figure.} \label{convergence}
\end{figure}

\begin{figure}
\includegraphics[width=8.5cm]{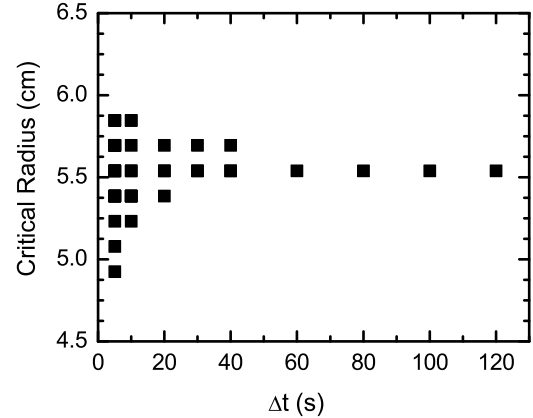}
\caption{Illustration of the convergence of the measured value of
$r_c$ as a function of the time interval over which the average
velocity is computed is increased.} \label{convergence2}
\end{figure}

\begin{figure}
\includegraphics[width=8.5cm]{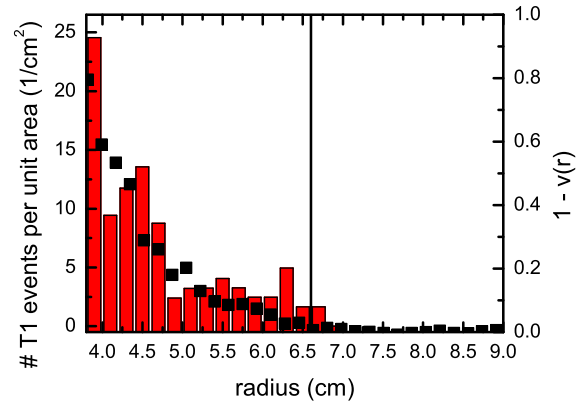}
\caption{(color online) The T1 event rate for $R = 9\ {\rm cm}$
and $\Omega = 0.11\ {\rm s^{-1}}$ is shown as a function of $r$
(red histogram, left hand axis). This is compared to $1 - v(r)$,
where $v(r) = v_{\theta}(r)/\Omega r$ is the scaled velocity
(black points, plotted against the right hand axis).} \label{T1}
\end{figure}

Given the possible breakdown of the continuum approximation, it is worth
testing the short time behavior and determining the approach to
steady state flow. This was done by considering a range of time
intervals over which to compute the average bubble displacements,
and the corresponding average velocity profiles. For short enough
time intervals, we were able to compute a histogram of the
probability distribution for $r_c$. Here, the probability
distribution is computed as follows. For each independent time
interval, an average velocity profile is computed. As reported in
Ref.~\cite{D04}, these profiles tend to be highly nonlinear, but
there is a well-defined radius at which the profile deviates from
a rigid body rotation. This point is taken as the critical radius
for that realization of the velocity profile. This is computed for
each independent segment of data from a single run, and the
probability distribution is generated from this set of data.

Figure~6 presents the probability distribution for the case of
$\Omega = 0.07\ {\rm s^{-1}}$, and for four different time
intervals used to compute the velocity profiles. As expected, the
longer the time interval, the narrower the distribution. However,
even for relatively short time intervals, the full width of the
distribution is only on the order of $1\ {\rm cm}$, or about $3\
{\rm bubbles}$. As one increases the averaging time, the mean of
the distribution remains constant to within 1~\%. Another way of
considering the approach to steady state is illustrated in Fig.~7.
This shows the computed critical radius for different sets of data
using a fixed time interval. The plot illustrates that for time
intervals greater than 50~s, the computed value of $r_c$ does not
change and that different measurements of $r_c$ give the same
value. (Due to the length of the run, there are two points for the
$\Delta t = 60\ {\rm s}$ data, but one point for each of the
larger values of $\Delta t$.)

A final question of interest is the spatial distribution of the T1
events and the correlation between the velocity profile and the T1
events. This is shown in Fig.~8 for $\Omega = 0.11\ {\rm s^{-1}}$
and $R = 9\ {\rm cm}$. Plotted here is a histogram of the number
of T1 events in each radial bin (normalized by the area of the
bin). This is compared to a plot of $1 - v(r)$. The velocity is
plotted in this form to better correlate the T1 and velocity
distributions. The solid line indicates the position of $r_c$. One
observes that the distribution of T1 events is consistent with the
measurements of the distribution of $r_c$ for short time averages.
As expected, one observes T1 events for values of $r$ close to,
but slightly greater than, $r_c$. However, there are no observed T1 events
for $r$ sufficiently greater than $r_c$.

\section{Discussion}

The main focus of the measurements in this paper is to test
carefully two aspects of the transition from solid to fluid
behavior in a slowly driven foam. First, whether or not the
transition is continuous or discontinuous in the rate of strain.
Second, whether or not there is a transition from a discrete to a
continuum flow regime. We used two standard models of complex
fluids to interpret the experimental results: the Herschel-Bulkley
model and a power-law model with a critical rate of strain. First,
we will discuss the evidence against the Herschel-Bulkley model.

As discussed in Sec.~III, the initial evidence against the
Herschel-Bulkley model is the fact that $r_c$ does not approach
$r_i$ as $\Omega$ goes to zero (see Fig.~2). Additional evidence
against the model is the results for $n$ presented in Fig.~3. The
fact that we essentially always measure an $n > 1$, in direct
contrast to the measurements of the stress as a function of rate
of strain, rules out the Herschel-Bulkley model. Finally, the
scaling of the velocity in Fig.~4 that strongly suggests the
existence of a discontinuity in the rate of strain is behavior
that is not possible in the context of the Herschel-Bulkley model.
One might be concerned that despite this evidence, one appears to
be able to fit the data to the Herschel-Bulkley model (as in
Fig.~1). However, given the discrete nature of the data, the
apparent fit to the Herschel-Bulkley model is most likely a result
of the non-physical values of $n$ that are obtained.

In contrast to the Herschel-Bulkley model, the fits to the
power-law model, and the scaling in Fig.~4a, strongly support the
applicability of the power-law model and the corresponding
discontinuity in the rate of strain. Therefore, it is reasonable
to state the the localized flow in the bubble raft is best
described by a coexistence of a power-law fluid and a solid
region, with a discontinuity in rate of strain at the coexistence
point. This has important implications for the jamming transition,
as it suggests the equivalent of a first order transition. Also,
it raises the important question of the mechanism that determines
the critical rate of strain. This will be the subject of future
work. One promising direction is to use a parallel shear cell. In
this case, one expects a uniform stress across the system, and the
global rate of strain is set by the boundaries. This can be used
to further test the nature of the critical rates of strain. For
example, in this case, it was determined to be $\dot{\gamma}_c =
0.07 \pm 0.01\ {\rm s^{-1}}$.

An interesting open question is which features of the flow are
determined by the details of the bubble raft. For example, how
does $\dot{\gamma}_c$ depend on the specifics of the solution used
to make the bubbles, the polydispersity, the bubble size, etc..
Likewise, we observed that the value of $n$ has potentially
significant variation. For example, even though $n < 1$ based on
fits to the power-law model (Fig.~3), it did differ from some past
measurements in bubble rafts \cite{LCD04}. This difference is
not surprising given that the details of the bubble rafts differed
to some degree in terms of the exact nature of the bubble size
distribution and the solutions used to make the bubbles.

The transition from a continuum regime to a discrete regime is
confirmed by the results presented in Figs.~4 and 5. Based on the
scaling arguments from Ref.~\cite{RBC05}, the data in Fig.~4a is
consistent with a single continuum model based on a power-law
fluid with a critical rate of strain for sufficiently high
rotation rates. The breakdown of the scaling indicates the
transition to the discrete regime. Directly plotting the critical
rate of strain for each rotation rate and system size, we find
that the transition between discrete and continuum behavior occurs
at $\Omega = 0.07 \pm 0.02\ {\rm s^{-1}}$. By considering two
system sizes, we were also able to establish that the behavior is
not dependent on system size. Conversely, we observe that the
systems we have considered are large enough to exhibit behavior
consistent with a continuum limit.

It is useful to compare our measurement of the transition to
discrete flow with the reported measurements in a
three-dimensional foam. Using the results in Fig.~2 for the
critical radius, a transition at $\Omega = 0.07 \pm 0.02 \ {\rm
s^{-1}}$ corresponds to a critical thickness for the flowing
regime of approximately 10 bubbles. For comparison, the same
transition is observed in three-dimensional bubbles for a
thickness of 25 bubbles and a rotation rate of $0.3\ {\rm s^{-1}}$
\cite{RBC05}. At this point, it would be useful to consider both
two and three dimensional models that capture this discontinuity
to determine the source of the differences in these measurements.
It might be dimensionality, but given the small differences, it is
more likely related to details of the bubbles, such as bubble
size, polydispersity, and surfactant composition. As current
simulations only predict continuous transitions, work is first
needed to elucidate the mechanism for the discontinuous
transition.

The existence of the discrete regime also has implications for the
jamming transition paradigm. The jamming transition is proposed in
terms of an externally applied stress. However, this stress is
often applied to the system using a constant external rate of
strain. The system is then assumed to pass through the jamming
transition as the stress increases as a function of total strain.
It is not clear how the jamming transition should be understood in
the discrete regime where the fluid-solid transition does not
appear to be well-described by continuum models. This will be an
interesting direction for future studies.

Finally, it is worth commenting on the insights gained by
consideration of the short time behavior of the system. We focused
on measurements of $r_c$ as a function of the averaging time (as
shown in Figs.~6 and 7). This was useful because it provided
confirmation that our results represent the steady-state of the
system. In addition, the comparison of $r_c$ and the spatial
distribution of T1 events are very suggestive.

The direct measurement of the distribution of T1 events confirmed
the solid-like behavior for $r > r_c$ in terms of the complete
absence of T1 events in the bulk of that region (Fig.~8). However,
the T1 events do extend beyond $r_c$ by a few bubble radii. This
is consistent with the width of the distribution of $r_c$ values
for sufficiently short time measurements (Fig.~6). As T1 events
occur right near $r_c$ (the steady-state average value), this will
generate short-time bubble motions that can extend up to a few
bubble diameters beyond $r_c$. These bubble motions can generate
deviations from rigid body motion. Therefore, the fluctuations in
$r_c$ have an important connection with the statistics of the T1
events at the transition from the fluid-like to the solid-like
behavior. It will be interesting in the future to compare this to
other fluctuating transitions between regions with different types
of flow such as those that have been observed in other complex
fluids \cite{SBMC03b,LAD06}.

\begin{acknowledgments}

This work was supported by a Department of Energy grant
DE-FG02-03ED46071. The authors thank K. Krishan and P. Coussot for
useful discussion.

\end{acknowledgments}

\section{Appendix}

In this Appendix, we present some of the details of the derivation
of the velocity profiles. The derivation of the profile for the
power-law fluid is standard \cite{BAH77}. One combines the the
general expression for the stress in a Couette geometry and the
relation between the stress and the rate of strain:
\begin{equation}
\sigma(r) = C/r^2 = \mu(\dot{\gamma}/\dot{\gamma}_c)^n,
\end{equation}
and combining all the constants, this simplifies to
\begin{equation}
\dot{\gamma} = D/r^{2/n}.
\end{equation}
Using the definition of $\dot{\gamma}$, one gets
\begin{equation}
\dot{\gamma} = r \frac{d}{dr}\frac{v_{\theta}(r)}{r} = D/r^{2/n},
\end{equation}
which can be directly integrated to get (dividing by $\Omega$)
\begin{equation} \label{eqn:PL4}
\frac{v_{\theta}(r)}{\Omega r} \equiv v(r) = \frac{A}{r^{2/n}} -
B.
\end{equation}
As given in Sec.~III, the boundary conditions determine $A$ and
$B$ in terms of $n$, $r_c$, and $r_i$.

The derivation for the  Herschel-Bulkley model follows along
similar lines, only now it is useful to explicitly write out the
constant in the stress relation:
\begin{equation}
\sigma(r) = C/r^2 = \sigma(r_i)r^2_i/r^2 = \tau_0 r^2_c/r^2.
\end{equation}
Notice, that there are two equivalent ways to write the constant
because it is based on the equivalence of torques in the radial
direction \cite{BAH77}. Again, we equate this expression for the
stress with the constitutive relation.
\begin{equation}
\tau_0 r^2_c/r^2 = \tau_0 + \mu \dot{\gamma}^n.
\end{equation}
This gives
\begin{equation}
\dot{\gamma} = (\tau_0/\mu)^{1/n}[r^2_c/r^2 - 1]^{1/n}.
\end{equation}
In this case, integration does not result in an analytic
expression, instead we get
\begin{equation} \label{eqn:HB2}
\frac{v_{\theta}}{r} = \Big(\frac{\tau_0}{\mu}\Big)^{1/n}
   \int_{r_i}^{r} \frac{1}{\rho}\Big(\Big(\frac{r_c}{\rho}\Big)^{2} - 1 \Big)^{1/n}
   d\,\rho + C.
\end{equation}
Here the constant $C = 0$ because $v_{\theta}/r = 0$ at $r = r_i$.
Requiring solid body rotation $v_{\theta}=\Omega r$ at $r=r_c$
gives
\begin{equation} \label{eqn:HB3}
\frac{v_{\theta}(r_c)}{\Omega r_c} = 1 =
\Big(\frac{\tau_0}{\mu}\Big)^{1/n}
   \int_{r_i}^{r_c} \frac{1}{\rho}\Big(\Big(\frac{r_c}{\rho}\Big)^{2} - 1 \Big)^{1/n}
   d\,\rho.
\end{equation}
Therefore, converting to $v(r) \equiv v_{\theta}(r)/(\Omega r)$,
\begin{equation} \label{eqn:HB4}
v(r) = \frac{1}{N}
   \int_{r_i}^{r} \frac{1}{\rho}\Big(\Big(\frac{r_c}{\rho}\Big)^{2} - 1 \Big)^{1/n}
   d\,\rho
\end{equation}
where
\begin{displaymath}
N = \int_{r_i}^{r_c}
\frac{1}{\rho}\Big(\Big(\frac{r_c}{\rho}\Big)^{2} - 1 \Big)^{1/n}
   d\,\rho.
\end{displaymath}
This is a useful form for numerically fitting the data by
approximating the integral.

%\bibliography{gsd06}

\end{document}